\documentclass{emulateapj}

\newcommand{\Fig}{Figure~}
\newcommand{\Tab}{Table~}
\newcommand{\NH}{N_{\rm H}}
\newcommand{\persqrcm}{{\rm cm}^{-2}}
\newcommand{\percubcm}{{\rm cm}^{-3}}
\newcommand{\kmpersec}{{\rm km~s}^{-1}}
\newcommand{\Kalpha}{{\rm K}_\alpha}
\newcommand{\Kbeta}{{\rm K}_\beta}
\newcommand{\logTK}{\log(T{\rm [K]})}
\newcommand{\source}{4U1820--303}

\slugcomment{to appear in ApJ}

\shorttitle{O and Ne absorption lines from NGC 6624}
\shortauthors{Futamoto et al.}

\begin{document}

\title{Detection of highly ionized O and Ne absorption lines in the X-ray spectrum of \source~ in the globular cluster, NGC 6624}
\author{Kazuo Futamoto, Kazuhisa Mitsuda, Yoh Takei, Ryuichi Fujimoto, and Noriko Yamasaki}
\affil{Institute of Space and Astronautical Science, Japan Aerospace Exploration Agency, \\
3-1-1, Yoshinodai, Sagamihara, Kanagawa, 229-8510, Japan}
\email{futamoto@astro.isas.jaxa.jp}

\begin{abstract}
We searched for absorption lines of highly ionized O and Ne in the
energy spectra of two Low-mass X-ray binaries, \source~ in the
globular cluster NGC6624 and Cyg X-2, observed with the {\it Chandra}
LETG, and detected \ion{O}{7}, \ion{O}{8} and \ion{Ne}{9} absorption
lines for \source.  The equivalent width of the \ion{O}{7} $\Kalpha$
line was $1.19^{+0.47}_{-0.30}$ eV (90 \% errors) and the significance
was 6.5 $\sigma$.  Absorption lines were not detected for Cyg X-2 with
a 90 \% upper limit on the equivalent width of 1.06 eV for \ion{O}{7}
$\Kalpha$.  The intrinsic line width was not resolved and an upper
limit corresponding to a velocity dispersion of $b = 420 ~\kmpersec$
was obtained for the \ion{O}{7} $\Kalpha$ line of \source. The ion
column densities were estimated from the curve of growth analysis
assuming several different values of $b$.  The absorption lines
observed in \source~ are likely due to hot interstellar medium,
because O will be fully photo-ionized if the absorbing column is
located close to the binary system.  The velocity dispersion is
restricted to $b = 200 - 420 ~\kmpersec$ from consistency between
\ion{O}{7} $\Kalpha$ and $\Kbeta$ lines, Ne/O abundance ratio, and H
column density.  The average temperature and the \ion{O}{7} density
are respectively estimated to be $\logTK ~= 6.2 - 6.3$ and $n_{\rm
OVII} = (0.7 - 2.3) \times 10^{-6} ~\percubcm$.  The difference of
\ion{O}{7} column densities for the two sources may be connected to
the enhancement of the soft X-ray background (SXB) towards the
Galactic bulge region.  Using the polytrope model of hot gas to
account for the SXB we corrected for the density gradient and
estimated the midplane \ion{O}{7} density at the solar neighborhood.
The scale height of hot gas is then estimated using the AGN absorption
lines.  It is suggested that a significant portion of both the AGN
absorption lines and the high-latitude SXB emission lines can be
explained by the hot gas in our Galaxy.
\end{abstract}


\section{Introduction}
The existence of hot ($T \sim 10^{5-6}$ K) interstellar medium in our
Galaxy has been known since the early 1970's, mainly from two kinds of
observations: the soft X-ray background (SXB) in the 0.1 -- 1 keV
range (e.g. \cite{Tanaka_Bleeker_1977}) and UV OVI absorption lines
(e.g. \cite{Jenkins_1978}) in OB stars.

The SXB in the so-called 1/4 keV band is dominated by the emission 
from the local bubble with $\logTK \sim 6$ around the sun \citep{Snowden_etal_1990}.
On the other hand,  in the 3/4 keV band, where the emission is dominated 
by \ion{O}{7} and \ion{O}{8} K lines, 
the majority of emission comes from hot interstellar medium 
with $\logTK \sim 6.2 - 6.4$, which is considered to widely distributed in
the Galactic disk, the bulge, and the halo
\citep[and reference therein]{Kuntz_Snowden_2000}.
  The SXB shows  enhancement in a circular region  $\sim 40 ^\circ$ 
in radius centered at the Galactic center.
  These are attributed to hot gas in the bulge of our Galaxy.
  Near the Galactic plane ($|b|~ ^{<}_{\sim}~ 10^\circ$), 
the emission is strongly absorbed by neutral matter.
  However, there remains emission comparable to that at  high latitude in the 3/4 keV band \citep{Snowden_etal_1997,Park_etal_1998,Almy_etal_2000}.
\cite{McCammon_etal_2002} clearly resolved for the first time the \ion{O}{7}, 
\ion{O}{8}, and a few other emission lines in the SXB at high latitude 
($b \sim 60 ^\circ$) with a rocket-borne microcalorimeter experiment.
   From the result they estimate that at least 42 \%  
of the soft X-ray background in the energy band 
that includes the O emission lines comes from thermal emission at $z<0.01$ 
and 38 \% from unresolved AGN.
  The origin of the remaining 20 \% (34\% for a 2 $\sigma$ upper limit) 
is still unknown and could  be extragalactic diffuse emission. 

  The \ion{O}{6} absorption line is considered to represent hot gas 
at lower temperatures, typically $\log T({\rm K}) = 5.5$.
  The \ion{O}{6} absorption lines observed in 100 extragalactic objects 
and two halo stars by {\it FUSE} are consistent with a picture
 where the hot gas responsible for the absorption has a patchy distribution 
but on the average has a plane-parallel exponential distribution 
with an average \ion{O}{6} midplane density 
of $1.7 \times 10^{-8} ~{\rm cm}^{-3}$ 
and a scale height of $\sim 2.3$ kpc \citep{Savage_etal_2003}.
   The average velocity dispersion of \ion{O}{6} absorption lines is $b=60~ \kmpersec$ with a standard deviation of $15 ~\kmpersec$.

On the other hand, \ion{O}{7}, \ion{O}{8}, \ion{Ne}{9},
and \ion{C}{6} absorption lines were detected in the energy spectra 
of bright active galactic nuclei (AGN) observed 
with the dispersive spectrometers on board 
the {\it Chandra} and {\it XMM-Newton} observatories.
 \citep{Nicastro_etal_2002,Fang_etal_2002,Rasmussen_etal_2003}.
  In spite of the fact that the redshift is consistent with 0,
 a large fraction of the absorption is considered to arise 
from the hot ($\logTK \sim 6.4$) plasma outside our Galaxy,
 i.e. the so-called warm-hot intergalactic medium 
(WHIM, \cite{Cen_Ostriker_1999,Dave_etal_2001}).
  For example, \cite{Rasmussen_etal_2003} argued 
that the scale height of the hot gas should be larger than 140 kpc 
in order to consistently explain the equivalent width of the \ion{O}{7} 
 absorption line and the intensity of the emission line at the same time.
 However, obviously the hot gas in our Galaxy contributes 
to the absorption lines to some extent.  


In order to constrain the density and distribution of hot gas 
which is responsible for \ion{O}{7} and \ion{O}{8} emission in our Galaxy,
 and  to constrain its contribution to the SXB and to the AGN absorption lines,
 we searched for absorption lines of highly ionized O in the energy spectra
 of Galactic X-ray sources.

In this paper we report the first detection of \ion{O}{7}, \ion{O}{8}
 and \ion{Ne}{9} absorption lines in the X-ray spectrum of \source~
 in the globular cluster NGC6624  observed
 with the low energy transmission grating (LETG) on board
 the {\it Chandra} observatory.
  We consider that the absorption lines are most likely
 due to the interstellar medium.
 We constrain the temperature and density of the plasma responsible
 for the absorption and compare the results with the models of the SXB, \
 the AGN absorption lines,
 and the SXB emission lines. 

Throughout this paper, we quote single parameter errors
at the 90~\% confidence level unless otherwise specified.

\section{Analysis and results}

In order to detect interstellar \ion{O}{7} and \ion{O}{8} absorption lines,
 we need to select an appropriate X-ray lighthouse for the study.
  First, the absorption by neutral matter must be small.
  We set a criterion, $\NH~^{<}_{\sim} ~2 \times 10^{21} ~\persqrcm$,
 so that the X-ray transmission at O line energies is larger than 0.2.
  On the other hand, a larger ionized O column density is expected
 for distant sources, and thus for sources with a large $\NH$.
Thus, we should select sources close to $\NH \sim 2 \times 10^{21} ~\persqrcm$.
 Sources at a high Galactic latitude may be better
 because a larger column-density ratio of ionized O to neutral H is expected.
 Finally, we need a bright X-ray source with a featureless X-ray spectrum.
 According to those criteria, we selected two low mass X-ray binaries,
 \source~ and Cyg X-2.  

In this section we first describe the analysis and results for
 \source~ and Cyg X-2.
 The analysis of the latter will be described only briefly
 because the analysis of the spectrum near O absorption of the same data set
 has already been described in detail in a separate paper
 \citep{Takei_etal_2002}. 

\subsection{\source}

\source~ is a Low-Mass X-ray Binary (LMXB) located in the globular cluster
 NGC 6624 at ($l$, $b$) = ($2.8^\circ$, $-7.9^\circ$).
  The distance of the star cluster is determined from the optical reddening,
 and the brightness of horizontal-branch and main-sequence turn-off stars
 \citep{Richtler_etal_1994,Kuulkers_etal_2003}.
   In this paper we adopt the value $7.6 \pm 0.4$ kpc
 from \citet{Kuulkers_etal_2003}.
 The authors also state that the peak brightness of X-ray bursts
 with photospheric expansion is consistent with this distance
 and the He Eddington limit. 
Adopting the distance of 7.6 kpc,  \source~ is located
1.0~kpc above the Galactic disk.  The neutral hydrogen column density
towards  the source is estimated to be  $\NH = N_{\rm HI}+2N_{\rm H_2} =1.9 \times 10^{21} ~\persqrcm$  from E(B-V)$=0.32 \pm 0.03$ \citep{Kuulkers_etal_2003,Bohlin_etal_1978}.  This is consistent with the total \ion{H}{1} column density
of our Galaxy, $N_{\rm HI} =1.5 \times 10^{21} ~\persqrcm$  from 21 cm radio observations \citep{Dickey_Lockman_1990}.

\subsubsection{Data reduction}

\source~ was observed with the LETG/HRC(High Resolution Camera)-S
 for 15.1~ks on March 10, 2000  (obsID 98).
  We retrieved the archival data from the CXC (Chandra X-ray Center).
  We used the data reprocessed by CXC on January 1, 2002.
  The ASCDS version number is 6.5.1. 
Throughout the present analysis,
 we used the software  CIAO 2.3 with the calibration data in CALDB 2.21.  

During the present observation, the source exhibited only 2.1 \% rms
 intensity variation on time scales longer than 10 minutes.
  Thus, we integrated all the data after the standard data screening.
We then summed the spectra of the positive and negative orders,
 and rebinned to 0.025~\AA~ bins, and subtracted the background.
The background was only $\sim 2~\%$ of the source counts in the wavelength
 region we used in the later analysis. 
The spectrum we obtained contains photons not only of the first 
order dispersion but also of the higher orders.
  We subtracted the higher order spectra using the method
described in \cite{Paerels_etal_2001} and \cite{Takei_etal_2002}.  

We estimated the statistical errors of the first order spectrum
by error propagation.
Because of the error propagation, the statistics of the spectral bins
are no longer completely independent. However, the increase of the
statistical errors after the higher order subtraction is smaller than
1\% in the wavelength ranges we use in the following analysis. We thus
treat them as independent in the spectral fitting.

According to the LETGS calibration report at the CXC
\footnote{see http://cxc.harvard.edu/cal/Letg/calstatus.html},
the wavelength calibration of LETGS spectra is accurate to 0.02~\%.
This is smaller by an order of magnitude than the statistical errors
of wavelength determined in the following analyses.

\subsubsection{O and Ne absorption lines}

\begin{figure*}
\includegraphics[width=0.33\textwidth]{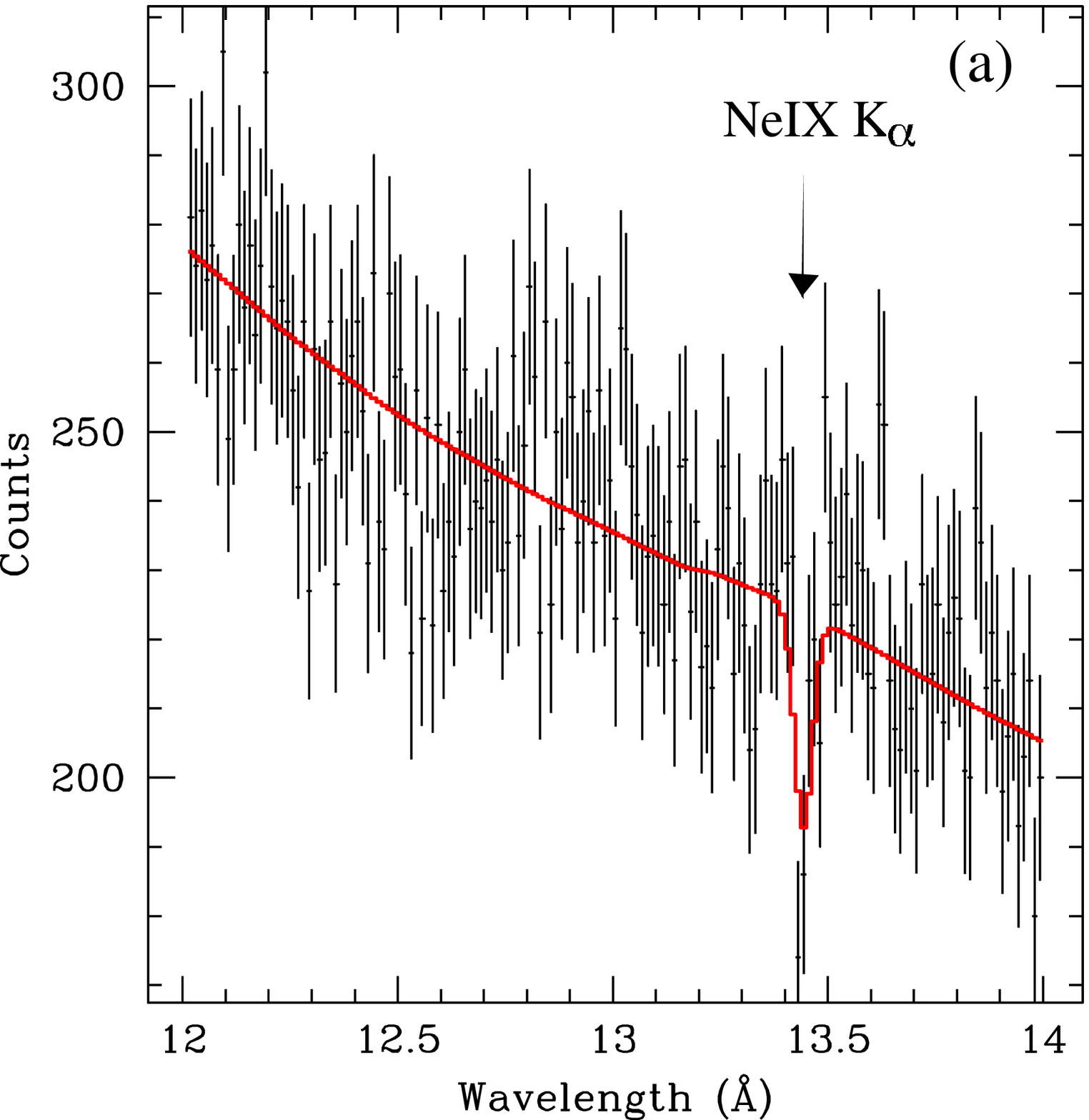}
\includegraphics[width=0.33\textwidth]{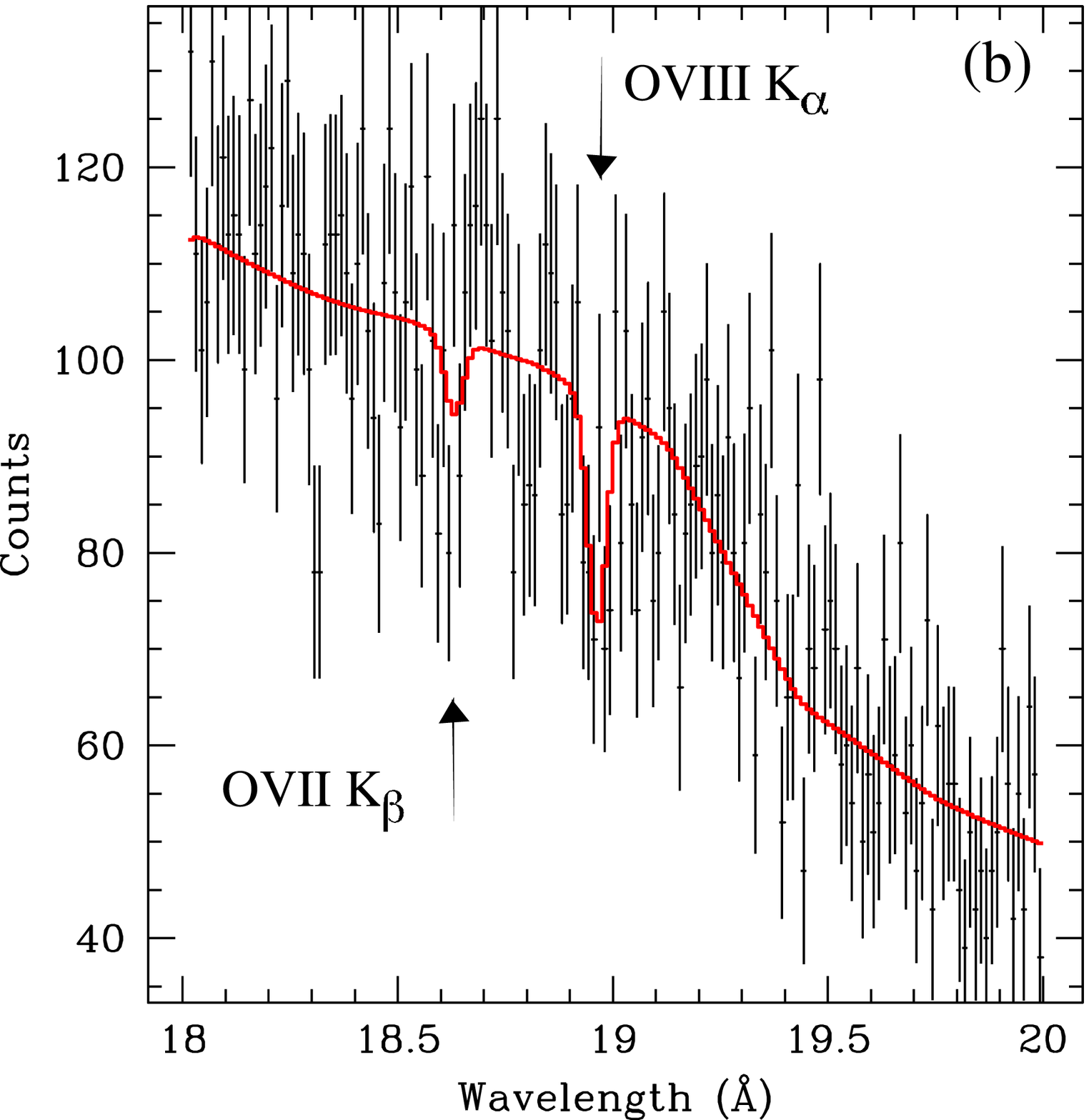}
\includegraphics[width=0.33\textwidth]{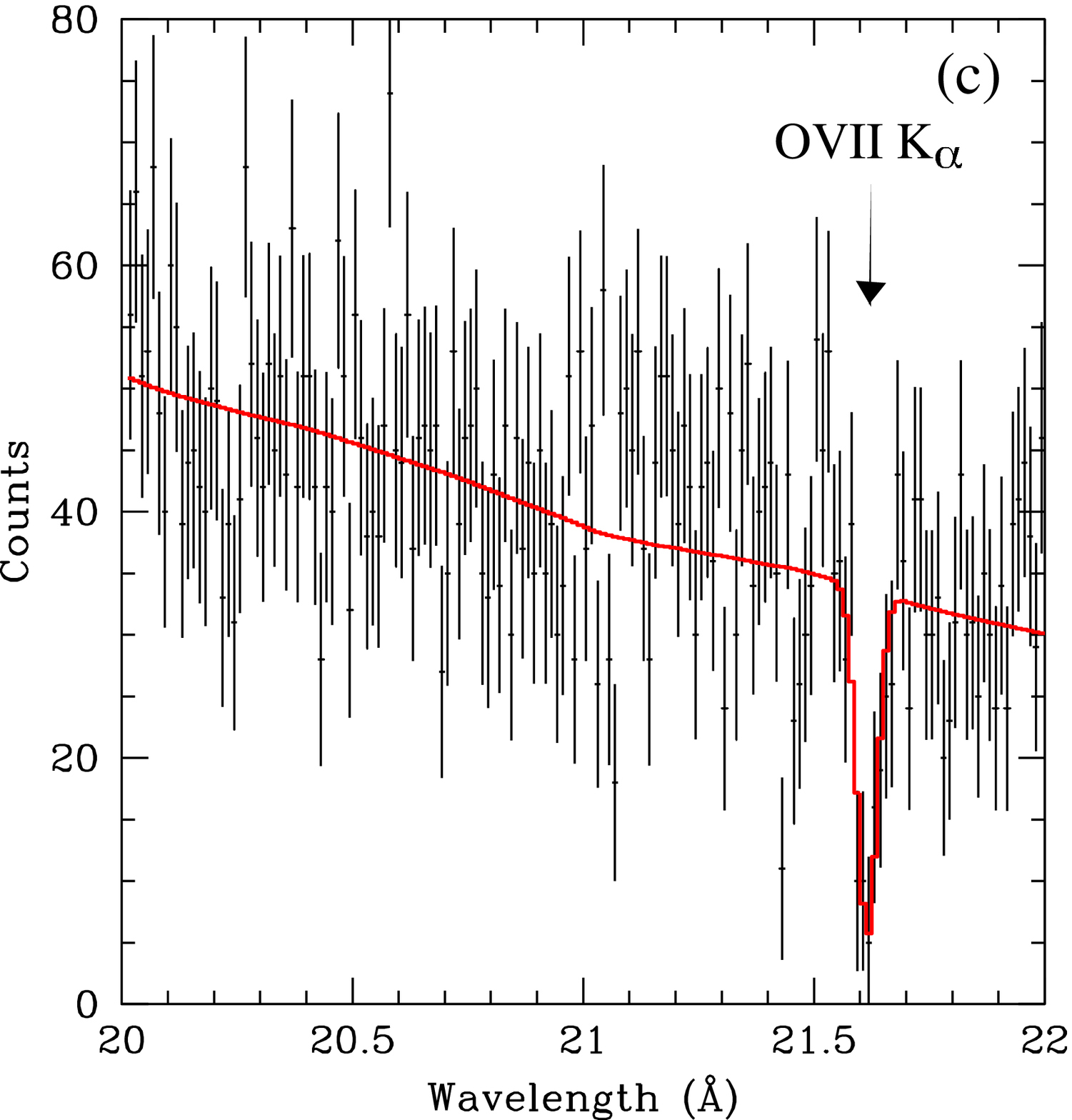}
\caption{Observed spectrum of \source~ in the wavelength ranges used for spectral fits for (a) \ion{Ne}{9},  (b) \ion{O}{8}, and  (c) \ion{O}{7} $\Kalpha$ absorption lines (data points with $1-\sigma$ vertical error bars) and  best fit model functions convolved with the telescope/detector response functions.   
}
\label{fig:OVII_Ka}
\end{figure*}

In the observed spectrum of \source, we clearly found an absorption
line at  $\lambda\sim$~21.6~\AA\  (see \Fig\ref{fig:OVII_Ka}(c)) in addition to the neutral-O absorption structures near $\lambda\sim$~23~\AA.    
From the wavelength, the absorption line can be identified as the \ion{O}{7} K$_\alpha$ resonant line.  We also noticed absorption lines corresponding to \ion{O}{8} K$_\alpha$ and \ion{Ne}{9} K$_\alpha$  (\Fig\ref{fig:OVII_Ka}(a)(b)).  We also looked at the wavelength corresponding to \ion{O}{7} K$_\beta$, but the absorption line was not clear.

In order to estimate the centroid wavelength, the intrinsic width, and the equivalent width of the lines, we performed  spectral fits.  
Since we fit a spectrum of relatively narrow wavelength range, we assumed a power-law function with absorption by neutral matter for the continuum spectrum.
We determined the continuum model parameters for Ne (\ion{Ne}{9}) 
 and O (\ion{O}{7} K$_\alpha$,  \ion{O}{8} K$_\alpha$,
 and  \ion{O}{7} K$_\beta$ lines) separately.
  The wavelength ranges are respectively 18 to 28 \AA~ for O
 and 12 to  14 \AA~ for Ne, excluding the narrow wavelength ranges of the 
absorption lines: 18.4 -- 19.2 \AA, 21.4 -- 21.8 \AA, and 13.3 -- 13.6 \AA.  
We represented the absorption with the {\tt tbabs} model
 in the `sherpa' program \citep{Wilms_etal_2000} with the H column density
 fixed to $1.9 \times 10^{21} ~\persqrcm$ for Ne.
  For O, we used the model employed by \cite{Takei_etal_2002} for Cyg  X-2.
  We set the O abundance of the {\tt tbabs} model to zero
 and represented the O absorption features
 with three edge models and two absorption lines. 

We then determined the absorption line parameters,
 fitting the spectra in the wavelength ranges 
shown in \Tab \ref{tab:fit_results}.
  In these fits, we fixed  the power-law index and the absorption parameters
 of the continuum to the values obtained in the previous fits.
 Thus, the free parameters are:  the normalization of the continuum,
 the centroid wavelength, the intrinsic width,
 and the equivalent width of the absorption line.
  We modeled the absorption line by adding a negative Gaussian
 to the continuum.
 The best fit parameters of the absorption lines are summarized
 in \Tab\ref{tab:fit_results}.  
In the table we expressed the intrinsic width of the line
 with the $b$ parameter,  which is related to the Gaussian $\sigma$
 of the fitting model as $b/c=\sqrt{2} \sigma /\lambda$
 ($c$ is the velocity of light).
  Thus $b$ is related to the temperature as $b = \sqrt{2 k_{\rm B} T / m_{i}}$
 in the case of thermal Doppler broadening
 where $k_{\rm B}, T, m_{i}$ are respectively the Boltzmann constant,
 the temperature, and the mass of the ion.  

As shown in \Tab\ref{tab:fit_results}, we obtained only an upper limit
 for the intrinsic width $b$.
  The present model gives ``negative X-ray intensity'' in the model photon
 spectrum when the intrinsic width is narrower than a certain value.
  However, even in such cases, the model spectrum convolved
 with the detector response function is positive and the equivalent width
 of the line is correctly estimated.
 We confirmed this by simulations, generating model spectra
 using an absorption line with a Voigt profile and fitting them
 with a negative Gaussian function.
  The difference between the model equivalent width and the best-fit value
 was less than 3\%.
  The column densities of ions will be estimated from the equivalent width
 utilizing the curve-of-growth analysis later.  

The centroid wavelengths of three lines are consistent
 with  \ion{O}{7} K$_\alpha$,  \ion{O}{8} K$_\alpha$,
 and \ion{Ne}{9} K$_\alpha$ at zero redshift, respectively.
  Those lines are statistically significant at the 6.5, 3.1,
 and 4.5 sigma levels.
  We obtained only an upper limit for the equivalent width
 of the \ion{O}{7} K$_\beta$ line.

\begin{table}[htb]
 \tiny

 \caption{Results of spectral fits of \ion{O}{7}, \ion{O}{8} and \ion
 {Ne}{9} absorption line features }
 \label{tab:fit_results}
 \begin{tabular}{lccccc} \hline \hline
          & fitting range & centroid $\lambda$   & $cz$ & Width ($b$) 
 & EW \\
   Line ID & (\AA) &  (\AA) &  ($\kmpersec$) 
        & ($\kmpersec$) & (eV) \\ \hline
 \source\\ 
   ${\rm O_{VII} K \alpha}$ 
       & 20-- 22 
 	  & $21.612^{+0.011}_{-0.006}$ 
       & -79 to +150
 	  & $< 420 $ 
 	  & $1.19^{+0.47}_{-0.30}$ \\
   ${\rm O_{VIII} K \alpha}$ 
       & 18-- 20 
 	  & $18.962^{+0.021}_{-0.015}$ 
       & -230 to + 330
 	  & (1) 
 	  & $0.54^{+0.23}_{-0.25}$ \\
   ${\rm O_{VII} K \beta}$ 
       & 18-- 20 
 	  & 18.629
 	  & 
 	  & (1) 
 	  & $< 0.48$ \\
   ${\rm Ne_{IX} K \alpha}$ 
       & 12-- 14 
 	  & $13.442^{+0.009}_{-0.016}$ 
       & -350 to +200
 	  & (1) 
 	  & $0.50 \pm 0.20$ \\
       \hline
 Cyg X-2\\
   ${\rm O_{VII} K \alpha}$ 
       & 20-- 22 
 	  & $21.602$ 
       & 
 	  & (1) 
 	  & $< 1.06$ \\
 	  \hline
 \end{tabular}
 (1) The error domain for O VII $\Kalpha$ line of \source~ is assumed  for the estimation of the error domain (or the upper limit) of the equivalent width. 
 \end{table}

\subsection{Cyg X-2}
Cyg X-2 is a LMXB located at ($l$, $b$)=($87.3^\circ$, $-11.3^\circ$).
  The distance is estimated to be 7.2 kpc,
 which locates the source at 1.4 kpc above the Galactic disk.
  The neutral hydrogen column density towards the source is $2.2 \times 10^{21} ~\persqrcm$ (see \cite{Takei_etal_2002} and references therein).    

The X-ray spectrum near the neutral O edge observed with the Chandra LETG/HRC-I
 was studied by \citet{Takei_etal_2002}.
  We used the spectrum and continuum model function in the paper
 and fitted a narrow wavelength range including the \ion{O}{7} absorption line.
  As shown in \Tab\ref{tab:fit_results}, we obtained only an upper limit
 for the \ion{O}{7} absorption line.

\subsection{Curve of growth and ion column densities}

\begin{figure*}
\plottwo{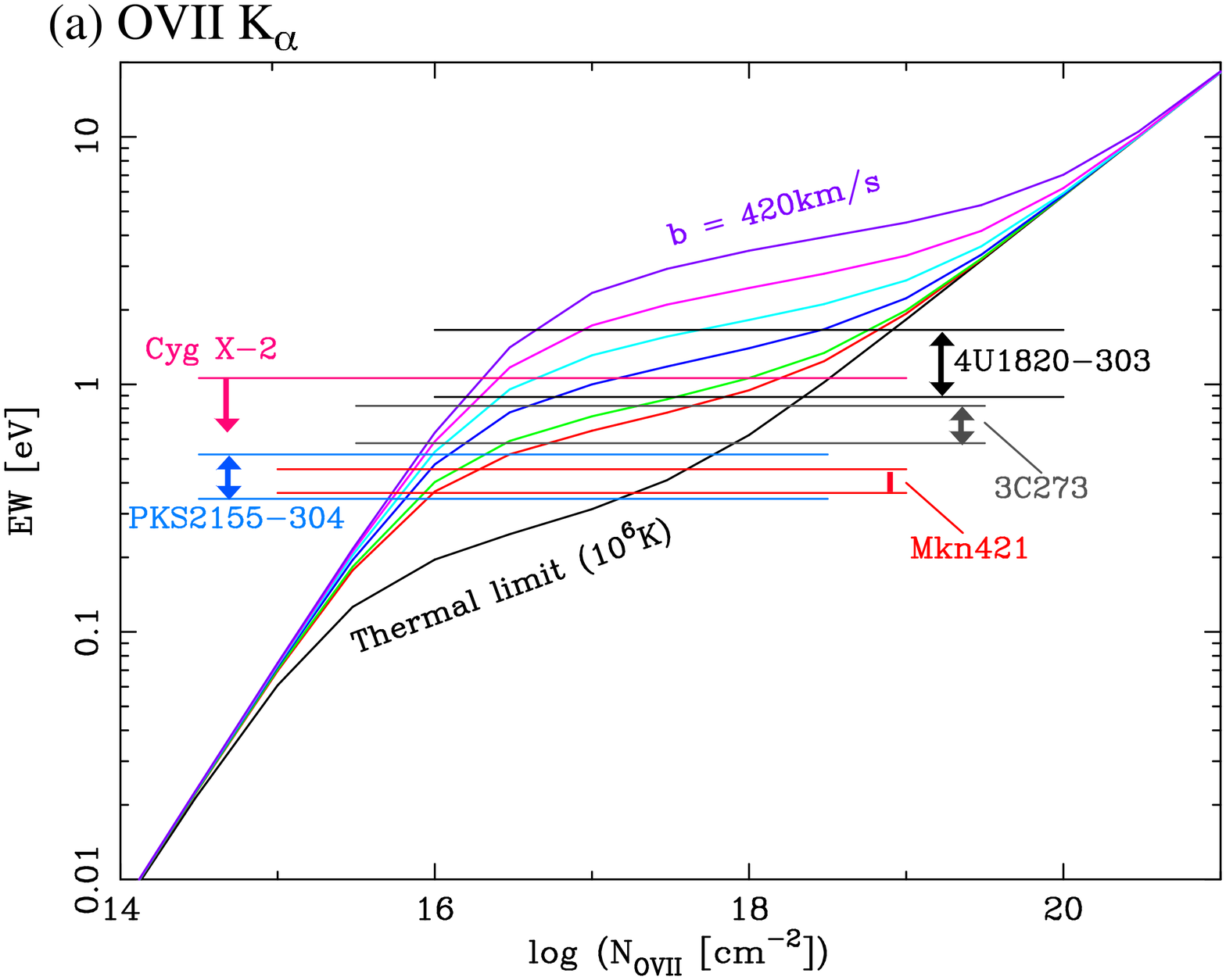}{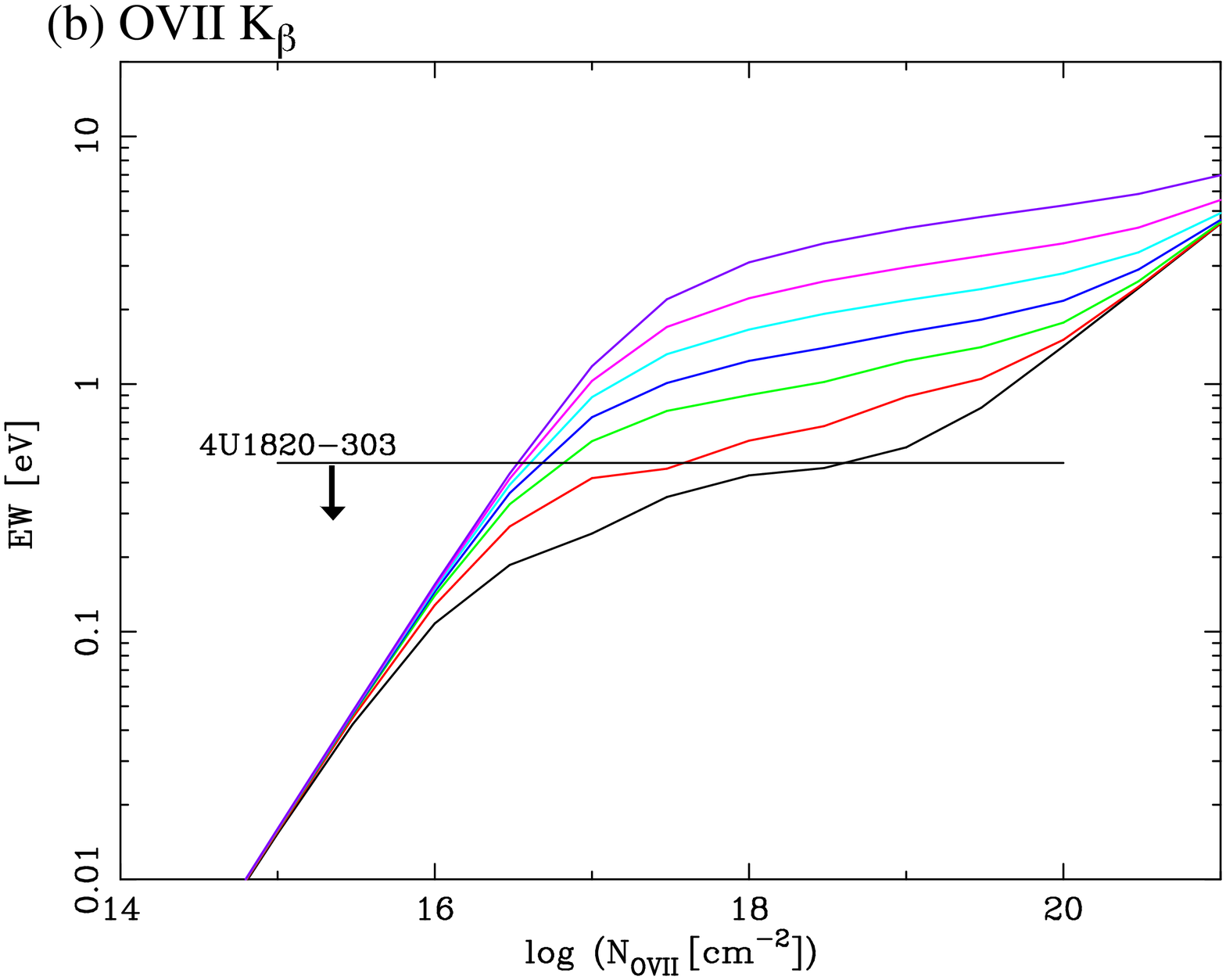}
\plottwo{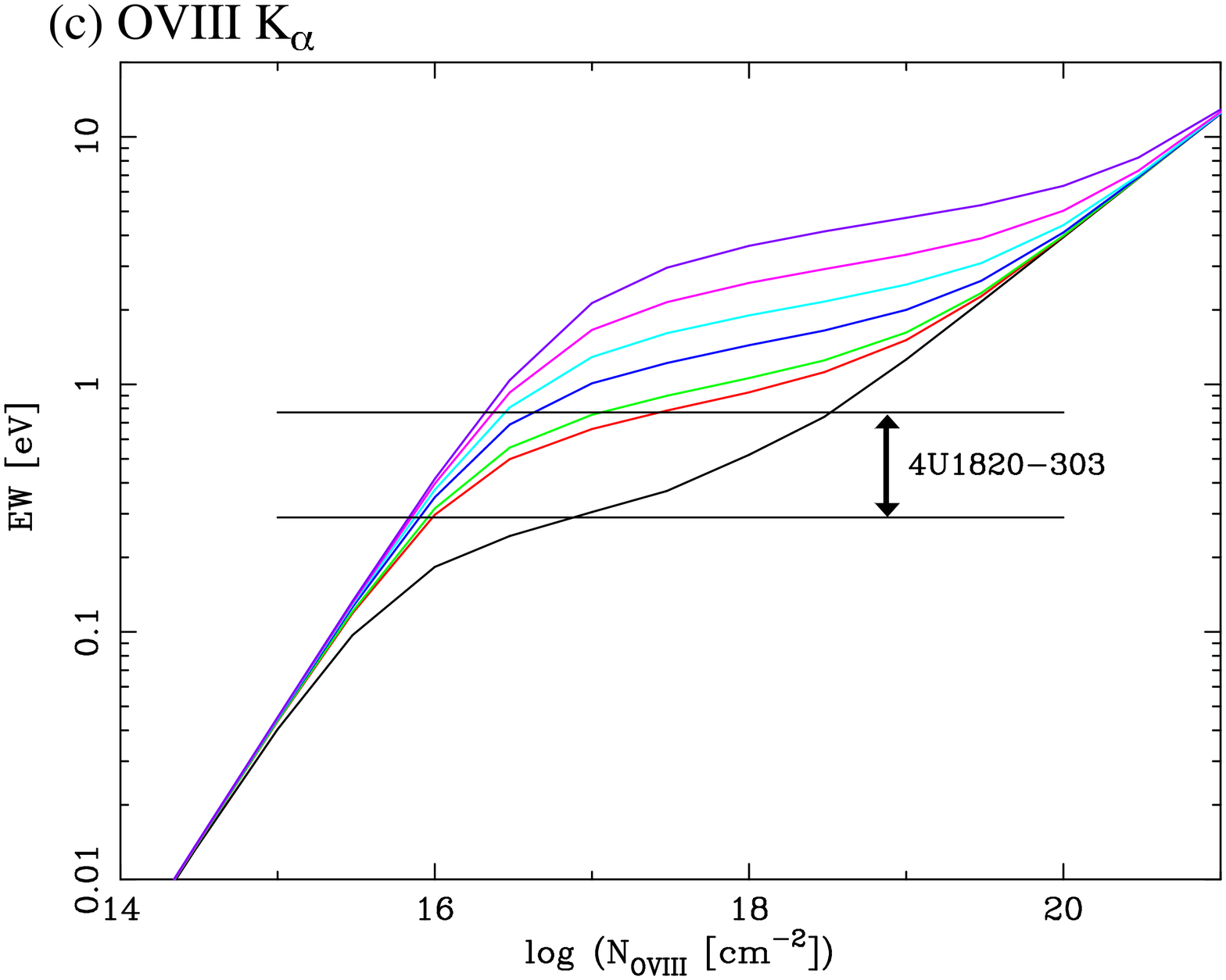}{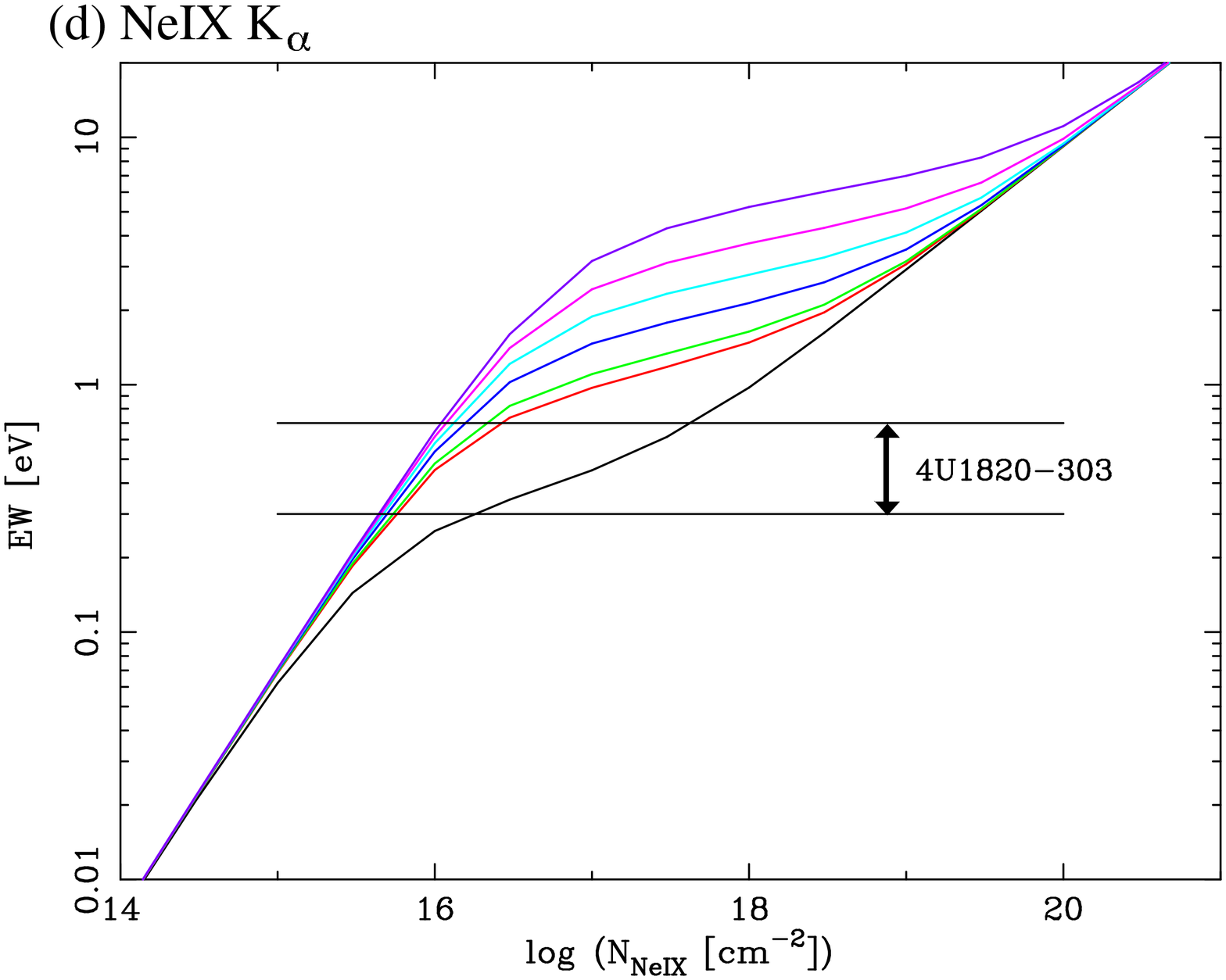}
\caption{
Curve of growth for (a) \ion{O}{7} $\Kalpha$, (b) \ion{O}{7} $\Kbeta$,
(c) \ion{O}{8} $\Kalpha$, and (d) \ion{Ne}{9} $\Kalpha$ absorption lines.  The different curves correspond to different values of velocity dispersion parameter, $b$ = 32 (for O) or  29 (for Ne), 60, 100, 140, 200, 280, and 420 $\kmpersec$. 
The minimum value of $b$ corresponds to thermal motion with $T = 10^{6}$ K,
and the maximum, $420 ~\kmpersec$, to the upper limit of the intrinsic
 line width of the 
\ion{O}{7} $\Kalpha$ line.  
The horizontal lines indicate the equivalent width constrained from observations.  In panel (a) (\ion{O}{7} $\Kalpha$), we show the equivalent width of AGN  from \cite{Rasmussen_etal_2003} in addition to \source~ and Cyg X-2.
}
\label{fig:cog}
\end{figure*}

\begin{figure}[t]
\plotone{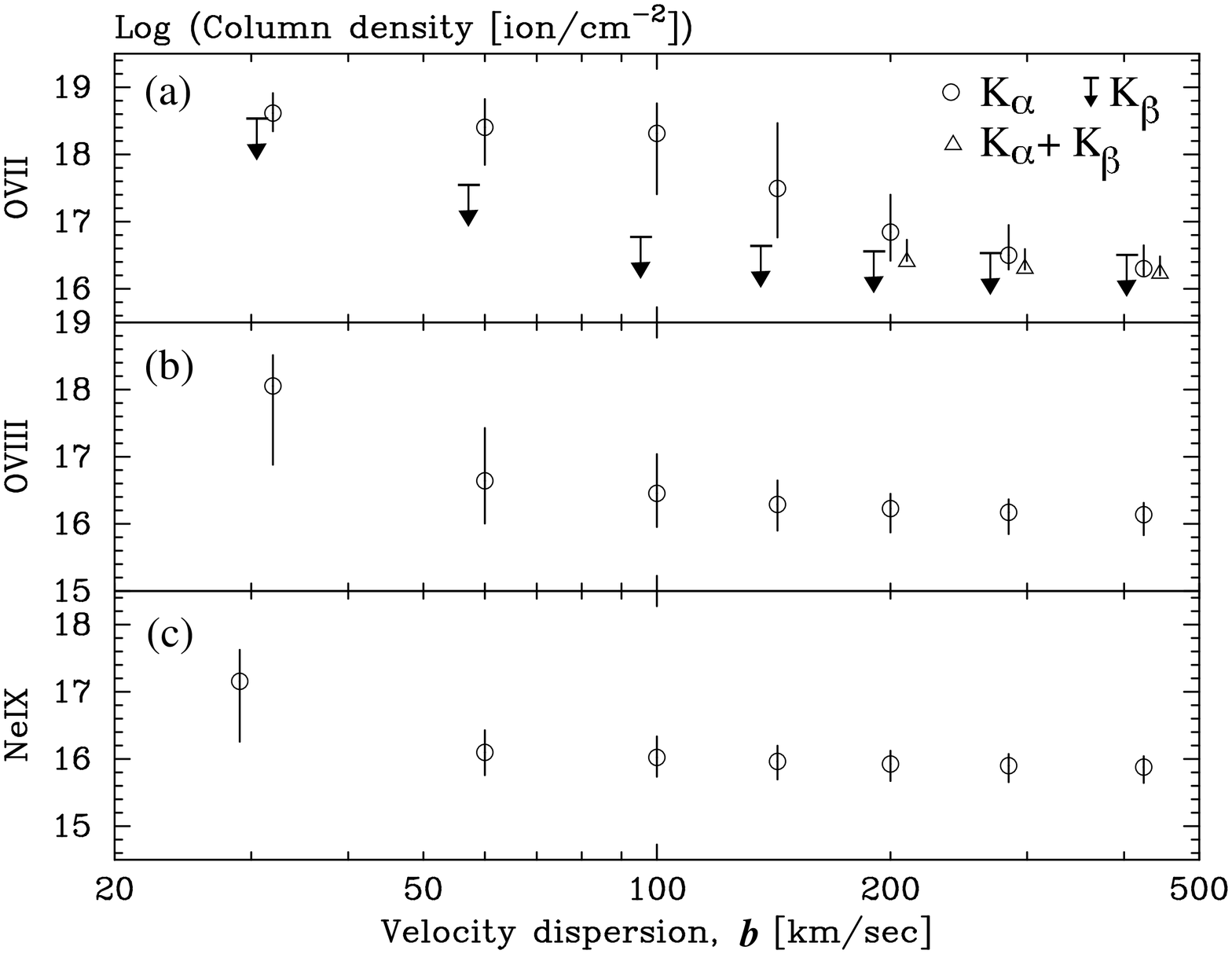}
\caption{
Column densities of (a) \ion{O}{7},  (b) \ion{O}{8}, and (c) \ion{Ne}{9} for \source~ as functions of assumed value of the velocity dispersion parameter $b$.  In panel (a), the column density obtained from the \ion{O}{7} $\Kalpha$ line (circles),  from \ion{O}{7} $\Kbeta$ lines (upper limits, data points shifted in $-b$ direction), and from combined fits of \ion{O}{7} $\Kalpha$ and $\Kbeta$ lines (triangles, only for $b \ge 200 \kmpersec$ and data points shifted in $+b$ direction) are respectively plotted.
}
\label{fig:column_density}
\end{figure}

In order to constrain the ion column densities,
we performed curve of growth analysis
(e.g. \cite{Nicastro_etal_1999,Kotani_etal_2000}).
 We adopted the oscillator strength and the transition probability
from \cite{Verner_etal_1996} for \ion{O}{7} $\Kalpha$, $\Kbeta$, 
and  \ion{O}{8} $\Kalpha$, and   from \cite{Behar_Netzer_2002}
for \ion{Ne}{9} $\Kalpha$.
From the analysis, we obtain the equivalent width of the line
as a function of the ion column density and the velocity dispersion. 
In \Fig\ref{fig:cog}, we show the curves of growth of the four ions
for several different values of $b$.
 As the maximum value of $b$, we adopted the upper limit
of the intrinsic width of the \ion{O}{7} K$_\alpha$ absorption line.
The lowest value of $b$ in \Fig\ref{fig:cog} corresponds
to thermal motion at a temperature of $1 \times 10^6$ K.
We show in \Fig \ref{fig:column_density} the \ion{O}{7}, \ion{O}{8}, and
the \ion{Ne}{9} column densities for \source~ as  functions of $b$.
  The error domain and the upper limit of the \ion{O}{7} column density ($N_{\rm OVII}$) obtained from $\Kalpha$ and $\Kbeta$ lines do not overlap with each other for $60 \ge b \ge 140 ~\kmpersec$, suggesting the velocity dispersion is either as small as the thermal velocity of $T \sim 10^6$ K or $b~ ^>_\sim ~200 ~\kmpersec$.  

We re-determined $N_{\rm OVII}$ for $b~ \ge ~200 ~\kmpersec$,
minimizing the sum of the two $\chi^2$ of the $\Kalpha$ and $\Kbeta$ 
spectral fits;
$$
\chi^2 = \chi^2_{\Kalpha}(EW_{\Kalpha}(N_{\rm OVII})) 
   + \chi^2_{\Kbeta}(EW_{\Kbeta}(N_{\rm OVII})),
$$
where all other spectral parameters were respectively optimized.
The best fit values are plotted in  \Fig \ref{fig:column_density} (a).
They are in the range, $\log(N_{\rm OVII}) = 16.20 - 16.73$. 
 
\section{Discussion}

\begin{figure}[t]
\plotone{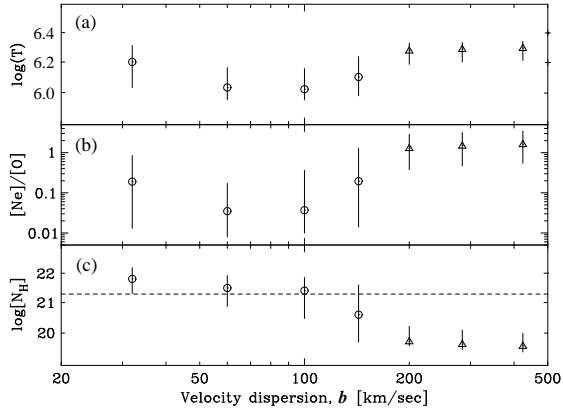}
\caption{
Temperature (a), Ne to O abundance ratio in the unit of solar ratio (b),
 and H column density (c) as functions of assumed value of the velocity
 dispersion parameter, $b$.  For $b \le 140 \kmpersec$, the \ion{O}{7}
 column density estimated from the $\Kalpha$ line is used to evaluate the plotted values, while for $b \ge 200 \kmpersec$, the combined analysis of $\Kalpha$ and $\Kbeta$ lines is used.  The horizontal broken line in panel (c) indicates the neutral H column density to \source. 
}
\label{fig:temperature}
\end{figure}

We searched for ionized O and Ne absorption lines in the energy
spectra of \source~ and Cyg X-2 observed with the {\it Chandra} LETG.
We detected \ion{O}{7} and \ion{O}{8} and \ion{Ne}{9} ${\rm K}_\alpha$
lines for \source~ and determined the equivalent widths and an upper
limit on the intrinsic width.  From the curve of growth, we estimated
the column densities as functions of the velocity dispersion, $b$.  In
order to make the equivalent widths of \ion{O}{7} $\Kalpha$ and
$\Kbeta$ consistent to each other, $b$ must be either as small as the
thermal velocity or $b~ ^>_{\sim} ~200 ~\kmpersec$. On the other hand,
we obtained only upper limits on absorption lines for Cyg X-2.  In
this section we will discuss the physical parameters of hot plasma
responsible for the \source~ absorption lines, and the possible
distribution of hot plasma in our Galaxy.

We first consider the possibility that the absorption lines observed
in \source~ are due to photo-ionized plasma associated with the binary
system.  In such cases, the temperature of the plasma can be lower
than $10^{6}$ K.  Thus, we calculated the curve of growth for zero
velocity dispersion to obtain the firm upper limit of the column
densities.  For \ion{O}{7}, it is $8 \times 10^{18} ~\persqrcm$.  From
the observed equivalent widths and the values of oscillator strengths,
we find that the column density of \ion{O}{7} is not smaller than that
of \ion{O}{8}.  Thus, it is natural to assume that \ion{O}{7} is the
dominant ionization state of O and that the ionization fraction of
\ion{O}{7} is on the order of $\sim 0.5$.  Then, assuming the solar
abundance, the hydrogen column density of the plasma must be $< 2
\times 10^{22} ~\persqrcm$.  \source~ is known to be an extremely
small system with an orbital period of 685 s.  The size of the binary
system is smaller than $0.1 R_\odot$ \citep{Stell_etal_1987}.  If the
plasma is located at a distance $r$ from the X-ray star, the
ionization parameter is $\xi = L_{\rm X}/(n r^2) \sim L_{\rm X}/(\NH
r)> 10^5$ with a luminosity of $2 \times 10^{37} {\rm erg~s}^{-1}$ and
$r < 10^{10}$ cm.  Then O will be fully photo-ionized
\citep{Kallman_McCray_1982}.  If the companion star is a He white
dwarf, the matter in the binary is He rich.  In that case, O should be
even more ionized because of the smaller number of electrons per
nucleon.

Since the above estimation is only valid for optically thin plasma
without an extra heat source, we ran the numerical simulation code,
CLOUDY96b4 \citep{Ferland_2002} to estimate the O ionization state for
more general cases.  We approximated the radiation spectrum from
\source~ with thermal bremsstrahlung of $kT=5$ keV.  We calculated for
the parameter ranges: $N_H = 2 \times 10^{22} - 2 \times 10^{24} ~
\persqrcm$, $r=10^{10}, 10^{11}, 10^{12}$ ~cm, and $T= 10^3, 10^4,
10^5, 10^6$ K, assuming solar abundance.  We also calculated for the
cases where 90 \% of the H atoms in the above absorption columns are
converted to He atoms. We found from the simulation that the observed
\ion{O}{7} and \ion{O}{8} column densities are reproduced only when
$r=10^{12}$ cm, $\NH = (1 -2 ) \times 10^{24} ~\persqrcm$, and $T \le
10^4$ K among all the combinations of parameters we tried.  Thus the
geometrical size of the plasma must be two orders of magnitude larger
than the binary size. It is not likely that the plasma responsible for
the absorption lines is related to the binary system.  We thus
consider it to be interstellar in the following discussion.

We can estimate the temperature of the hot medium from the ratio of
\ion{O}{7} to \ion{O}{8} column densities, assuming that these
absorption lines arise from the same plasma in collisional ionization
equilibrium.  In reality, the hot medium may be multiple plasmas of
different temperatures.  Thus, the temperature needs be regarded as an
"average" defined by the \ion{O}{7} - \ion{O}{8} ratio. In
\Fig\ref{fig:temperature} (a), the temperature is plotted as a
function of $b$.  For the ionization fraction we used the table in
\citet{Arnaud_Rothenflug_1985}.  As to the column density of
\ion{O}{7}, we used the value from the combined $\Kalpha$ and $\Kbeta$
fit for $b \ge 200 \kmpersec$, and the value from $\Kalpha$ for
others.
  The temperature is restricted to a relatively narrow range
 of $\logTK = 6.0 - 6.3$.
  The temperature range is consistent with the previous temperatures
 estimated for the SXB emitting plasma \citep{Kuntz_Snowden_2000}.

Given the temperature range, we can restrict the Ne to O abundance ratio
 by further assuming that the Ne is in collisional ionization
 equilibrium with O.
  We can also estimate the hydrogen column density of the hot plasma, $\NH$(hot),
 assuming solar abundance  
 \citep{Anders-Grevesse_1989} 
 for O.
  In \Fig\ref{fig:temperature} (b) and (c),
 we show the Ne/O abundance ratio in units of the solar ratio
 and $\NH$(hot) as functions of $b$.
In panel (c), we indicated the neutral $\NH$ with a horizontal broken line.
From these plots, we find that if $b < 140 ~\kmpersec$,
 $\NH$(hot) is comparable to or even larger than the $\NH$ of neutral medium
 and that the Ne/O abundance ratio must be significantly smaller
 than the solar value.
Since both are unlikely, we can exclude small $b$.
 Thus it is likely that $b~ ^>_{\sim} ~200 ~\kmpersec$,
 and we obtain the constraints: $\log\NH$(hot) = 19.41 -- 20.21,
 $\log N_{\rm OVII}$ = 16.20 -- 16.73, and $\logTK = 6.19 - 6.34$. 
Dividing the column densities by the distance,
 we obtain the densities averaged over the volume of the column,  
 $<n_{\rm H}{\rm (hot)}> = (1.1 - 7.0) \times 10^{-3} ~\percubcm$
 and $<n_{\rm OVII}> = (0.7 - 2.3) \times 10^{-6} ~\percubcm$, respectively.
 The hot plasma is likely to have a patchy distribution. Thus the local densities 
 are smaller than the volume average.  The volume filling factor  is  
 estimated to be $\sim 0.5$ in the solar vicinity,  although there are large uncertainties \citep{Mathis_2000} .
 
We did not detect an absorption line in Cyg X-2 in spite of the distance,
$\NH$, and the Galactic latitude which are all similar to those of \source.
Although the difference in the absorption line equivalent width
between Cyg X-2 and \source~ is not statistically very significant
because the upper limit of the \ion{O}{7} $\Kalpha$ equivalent width
of Cyg X-2 is higher than the lower boundary of the error domain
of \source, this may suggest that the average density of the hot plasma 
is lower for the line of sight of Cyg X-2.  
The SXB map observed with ROSAT shows enhancement in the circular region
of $\sim 40^\circ$ radius centered on the Galactic center
\citep{Snowden_etal_1997}.
This may explain the higher absorption column density for \source. 
 In order to quantitatively compare our results with the SXB emission, 
 we use the polytropic models of the hot gas  constructed  by \cite{Wang_1998}
and \cite{Almy_etal_2000} which reproduce the all sky SXB map. 
They assumed  nonrotating hot gas of polytropic index 5/3 
in  hydrostatic equilibrium with the Galactic potential 
\citep{Wolfire_etal_1995}.  
The model is described by
two parameters, the normalization factor of 
the polytrope $k$ and the pressure at the Galactic center $P_0$, or equivalently
the temperature $T_0$ and electron density $n_0$ at the Galactic center.  
\cite{Almy_etal_2000} included an additional isotropic component in the model 
SXB map to explain the surface brightness at high latitudes ($|b|~>~80^\circ$), 
but \cite{Wang_1998} did not.  
By fitting 
the ROSAT 3/4 keV all-sky map with the model,
\cite{Almy_etal_2000} obtained $k = P/\rho^{-5/3} = 1.45 \times 10^{32} 
~{\rm cm}^{4}~{\rm g}^{-2/3}~{\rm s}^{-2}$ and $P_0/k_{\rm B} = 1.8\times 10^5~ \percubcm~ {\rm K}$, 
which correspond to  $\log(T_0{\rm [K]}) = 6.92$ and  $n_0 = 1.1 \times 10^{-2}$.
Using these parameter values, we can calculate the temperature and density of the polytrope plasma at a given location in the Galaxy.  For example, at the solar neighborhood, $\logTK~ = ~6.20$ and
$n_{\rm H}{\rm (hot)} ~= ~8\times10^{-4} ~\percubcm$.
Comparing these values with the model parameters by \cite{Wang_1998},
 $\logTK = 6.23$ and $n_{\rm H}{\rm (hot)} =1.1 \times10^{-3} ~\percubcm$
 at the solar neighborhood, we consider there is at least a $\sim$ 30\% uncertainty in the model density, and the uncertainty is partly related to the interpretation of the high latitude emission. We adopt parameters  by \cite{Almy_etal_2000} in the following estimation. 

The model predicts $\NH({\rm hot}) = 3.9 \times 10^{19} ~\persqrcm$ for \source~ and  $\NH({\rm hot}) = 1.5 \times 10^{19} ~\persqrcm$ for Cyg X-2. However, assuming the solar abundance,
the \ion{O}{7} column density towards \source~ is expected to be $N_{\rm
OVII} = 0.7 \times 10^{16} ~\persqrcm$ which is $\sim 30$ \% smaller
than that of Cyg X-2 ($N_{\rm OVII} = 1.1 \times 10^{16} ~\persqrcm$).
  This is because the temperature increases towards \source~ along the
 line of sight up to $\logTK = 6.6$, while the \ion{O}{7} ionization
 fraction decreases very rapidly for $\logTK~ ^{>}_{\sim} ~6.4$ and the
 temperature stays in the range of $\logTK = 6.2 - 6.1$ along the line of
 sight of Cyg X-2.
 We thus need to modify the model to explain the column densities.  
The ratio of \ion{O}{7} column densities in two different directions is dependent on $T_0$ but not on $n_0$. 
\cite{Snowden_etal_1997} estimated the temperature of emission from the bulge region to be $\logTK = 6.6$ from the ratio of counts in the 3/4 keV and 1.5 keV bands.  Thus it may be possible to reduce $\log(T_0 {\rm[K]}) $ to this value.  Of course, we need to increase $n_0$ at the same time in order to reproduce the SXB intensity. 
The  ratio of \ion{O}{7} column densities to \source~ and to Cyg X-2 predicted by the polynomial model  increases when 
$\log(T_0 {\rm[K]}) $  is reduced from 6.9, and it is in the range of 1 -- 2.2 for  $\log(T_0 {\rm[K]}) = 6.7 - 6.6$. 
Thus the model can be made consistent with the present observations.
In this parameter range, the \ion{O}{7} density at the solar vicinity $<n_{\rm OVII, S}>$  is higher than the  average over the line of sight for \source.  The ratio of the two, i.e.  $r = <n_{\rm OVII, S}> /<n_{\rm OVII}> $
is in the range $0.77 - 0.50$  for $\log(T_0 {\rm[K]}) = 6.7 - 6.6$.
Using $r$, we can write   $<n_{\rm OVII, S}>  = (0.4 - 1.2) \times 10^{-6} (r/0.5) ~\percubcm$ from the observations.  
This is consistent with the previous estimate of the hot plasma density in the solar neighborhood \citep{Snowden_etal_1990}.

Now let us compare our results  with the AGN absorption lines and the high latitude SXB emission lines.   
The column densities of \ion{O}{7} for 3C 273, Mkn 421, and PKS 2155-304 are  respectively estimated  to be in the ranges of $\log(n_{\rm OVII}) =$ 16.0 --16.4 , 15.7 -- 15.9, 15.7 -- 16.0  assuming the velocity dispersion of $b = 200 - 420$ km/s and the equivalent widths from  \cite{Rasmussen_etal_2003}   (see \Fig \ref{fig:cog} (c)).  Then,  assuming  a vertical exponential distribution and a midplane density at the solar vicinity as estimated above, 
the scale height is estimated to be  $h = (2 - 20) \times (r/0.5)^{-1}$ kpc.   This suggests that a significant portion of the \ion{O}{7} absorption observed in the AGN spectra is of Galactic origin.   This scale height is consistent with the polytrope model, which predicts $h = 8$ kpc at the solar neighborhood.
Using the table in SPEX ver 1.10
(http://rhea.sron.nl/divisions/hea/spex),
 the intensity of OVII triplet lines in the direction of
 $(l, b) = (70^\circ, 60^\circ)$ is estimated to be
 $(2-20) (r/0.5)^{2}(h/10{\rm kpc})(1/f) ~{\rm photons} ~{\rm cm}^{-2} ~{\rm s}^{-1} ~{\rm str}^{-1}$, where $f$ is the volume filling factor of the hot gas.  This count rate  is  consistent with
 $4.8\pm 0.8~ {\rm photons}~ {\rm cm}^{-2} ~{\rm s}^{-1} ~{\rm str}^{-1}$, 
 as obtained by \cite{McCammon_etal_2002}.

Our estimate for the vertical scale height of hot gas is a factor of $^{>}_{\sim} 7$ smaller than the 140 kpc
 estimated by \cite{Rasmussen_etal_2003}.
The discrepancy is due to the difference in the abundance
 (1 solar v.s. 0.3 solar),
the spatial distribution ($\sim$~ exponential v.s. uniform),
and the temperature ($\logTK \sim $ 6.2 v.s. 6.4,
 resulting in a difference in the line emissivity of a factor of about 4). \\

In conclusion, the highly ionized O and Ne absorption lines observed
in \source~ are likely due to hot interstellar medium.
The velocity dispersion is restricted
to the range $b = 200 - 430 ~\kmpersec$, and 
the temperature 
$\logTK = 6.2 - 6.3$. The average densities along the line of sight 
 are $n_{\rm H}{\rm (hot)} = (1.1 - 7.0) \times 10^{-3} ~{\rm cm}^{-3}$
 and $n_{\rm OVII} = (0.7 - 2.3) \times 10^{-6} ~{\rm cm}^{-3}$, respectively.
  The higher \ion{O}{7} column density for \source~ than for Cyg X-2
 may be connected to the enhancement of the SXB towards
 the Galactic bulge region.
 Using the polytrope model
 which reproduces the all-sky SXB map, we  corrected for the density gradient
 along the line of  sight and estimated the midplane
 \ion{O}{7} density at the solar neighborhood. 
Combining this value with the absorption lines observed in the AGN we estimated the 
vertical scale height of hot gas to be in the range, 2 -- 20 kpc.
The intensity of the high latitude SXB emission lines is consistent with these estimations.
Thus we suggest that a significant fraction of both the AGN absorption lines and
the SXB emission lines can be explained  by hot gas in our Galaxy.

The present result demonstrates that the absorption lines
of Galactic X-ray sources are a powerful tool for constraining
the physical state of the hot interstellar medium.
A  search for absorption lines in other Galactic sources is urged to
further constrain the plasma distribution.
We consider that an observation of Cyg X-2 with a longer exposure time
is particularly important.

\acknowledgements We are very grateful to D. McCammon for valuable comments and discussions, and to D.Audley for careful review of the manuscript.  This work was supported in part by the Grants-in-Aid by MEXT/JSPS, Japan (KAKENHI 14204017, 12440067, and 15340088).

\end{document}